\documentclass[useAMS,fleqn,usenatbib]{mnras} 
\usepackage[T1]{fontenc}
\usepackage{ae,aecompl}

\usepackage{amsmath}	
\usepackage{amssymb}	
\usepackage{graphics,graphicx,epsfig}
\usepackage{mathrsfs}
\usepackage{amssymb,amsmath,amsfonts,amssymb,subfigure}

\newcommand{\cmg}{{\rm cm^2\,g^{-1}}}
\newcommand{\mstar}{M_{\star}}
\newcommand{\mhalo}{M_{\rm halo}}
\newcommand{\msun}{{\rm M}_{\odot}}
\newcommand{\pc}{\rm pc}

\title[SIDM on FIRE]{SIDM on FIRE: Hydrodynamical Self-Interacting Dark Matter simulations of low-mass dwarf galaxies}

\author[Victor H. Robles et al.]{Victor H. Robles$^{1}$\thanks{E-mail: roblessv@uci.edu(UCI)}, James S. Bullock$^{1}$,
Oliver D. Elbert$^{1}$, Alex Fitts$^{2}$,
\newauthor
Alejandro Gonz\'{a}lez-Samaniego$^{1}$, Michael Boylan-Kolchin$^{2}$, Philip F. Hopkins$^{3}$, 
\newauthor 
Claude-Andr\'{e} Faucher-Gigu\`{e}re$^{4}$, Du\v{s}an Kere\v{s}$^{5}$ and 
Christopher C. Hayward$^{3,6}$\\
$^{1}$Department of Physics and Astronomy, University of California, Irvine, CA 92697,USA\\
$^{2}$Department of Astronomy, The University of Texas at Austin, 2515 Speedway, Stop C1400, Austin, TX 78712-1205, USA\\
$^{3}$TAPIR, California Institute of Technology, Pasadena, CA, USA\\
$^{4}$Department of Physics and Astronomy and CIERA, Northwestern University, Evanston, IL, USA\\
$^{5}$Department of Physics, Center for Astrophysics and Space Sciences, University of California, San Diego, La Jolla, CA, USA\\
$^{6}$Center for Computational Astrophysics, Flatiron Institute, 162 Fifth Avenue, New York, NY 10010, USA}

\pubyear{2017}

\begin{document}
\label{firstpage}
\pagerange{\pageref{firstpage}--\pageref{lastpage}}
\maketitle

\begin{abstract}We compare a suite of four simulated dwarf galaxies formed in
  $10^{10} \msun$ haloes of collisionless Cold Dark Matter (CDM) with galaxies
  simulated in the same haloes with an identical galaxy formation model but a
  non-zero cross-section for dark matter self-interactions. These cosmological
  zoom-in simulations are part of the Feedback In Realistic Environments (FIRE)
  project and utilize the FIRE-2 model for hydrodynamics and galaxy formation
  physics.  We find the stellar masses of the galaxies formed in
  Self-Interacting Dark Matter (SIDM) with $\sigma/m= 1\, \cmg$ are very similar
  to those in CDM (spanning $\mstar \approx 10^{5.7 - 7.0} \msun$) and all runs
  lie on a similar stellar mass -- size relation.  The logarithmic dark matter
  density slope ($\alpha=d\log \rho / d\log r$) in the central $250-500 \, \pc$
  remains steeper than $\alpha= -0.8$ for the CDM-Hydro simulations with stellar
  mass $\mstar \sim 10^{6.6} M_{\odot}$ and core-like in the most massive galaxy.
  In contrast, every SIDM hydrodynamic simulation yields a flatter profile, with
  $\alpha >-0.4$.  Moreover, the central density profiles predicted in SIDM runs
  without baryons are similar to the SIDM runs that include FIRE-2 baryonic
  physics. Thus, SIDM appears to be much more robust to the inclusion of
  (potentially uncertain) baryonic physics than CDM on this mass scale,
  suggesting SIDM will be easier to falsify than CDM using low-mass
  galaxies. Our FIRE simulations predict that galaxies less massive than
  $\mstar \la 3 \times 10^6\,\msun$ provide potentially ideal targets for
  discriminating models, with SIDM producing substantial cores in such tiny
  galaxies and CDM producing cusps.
\end{abstract}




\section{Introduction}

The dark energy ($\Lambda$) + cold dark matter (CDM) model assumes the dark
matter is non-relativistic at decoupling and effectively collisionless, although
it is weakly interacting with the standard model of particles. $\Lambda$CDM is in
remarkable agreement with a variety of cosmological data on large scales
\citep{komatsu11,planck14}, but its consistency with observations on the scale
of dwarf galaxies is less clear. The predicted dense centers of CDM haloes are at
the root of two of the most notable issues: the \textit{cusp-core} problem
states that inner density profiles of dark-matter-dominated systems such as low
mass and low surface brightness (LSB) galaxies appear to be cored, contrary to
CDM-predicted cuspy centers
\citep{moore94,simon05,oh11,chan15,zhu16,kuz11,kuzspe11}; and the \textit{too
  big to fail} problem, which is that dark matter-only (DMO) simulations
predict a substantial population of massive, centrally-concentrated subhaloes
that does not appear to be present around the Milky Way (MW) or M31
\citep{boylan11,shealg14}.

These issues have driven substantial efforts to understand whether the
discrepancies between theory and observations lie in an incomplete modeling of
baryonic physics with the CDM paradigm. One particularly relevant prospect is
the realization that bursty star formation, with accompanying violent
gravitational potential fluctuations, may have the ability to re-shape the
central gravitational potentials of even dark-matter-dominated systems
\citep{gov10,gov12,pon12}. Subsequent papers have shown that bursty star
formation over an extended period can be effective in transforming a cusp to a
core and in reducing the central densities of the DM halo
\citep{ono15,chan15,read16,tollet16}; additionally, baryonic physics could also
help to alleviate the too big to fail problem \citep{chan15,wet16,adi12}.  The
results from existing CDM simulations of dwarf galaxies imply that variations in
the SFH of a galaxy have a large impact on the associated DM halo, even when
controlling for the host galaxy's stellar mass \citep{ono15}.

However, not all modern cosmological simulations of dwarf galaxies result in a
cored density distribution for dwarf galaxies. Smoother star formation histories
obtained via different assumptions for star formation (e.g., \citealt{saw16})
lead to cuspy profiles. Even simulations that do result in feedback-induced
cores typically find there is a limit to this process: as the halo mass
decreases, decreased star formation efficiency renders core creation (on the
scale of hundreds of parsecs) ineffective for galaxies with
$\mstar \la 10^6\,\msun$ \citep{chan15,tollet16,fitts16}.  Furthermore, properly
addressing the problems found in low-mass galaxies
($\mstar \leq10^9$M$_{\odot}$) requires high-resolution simulations that can
describe the central region of the dwarf DM haloes where these galaxies are
hosted.  Failure to resolve the dense centers of dwarfs can result in artificial
cores in the DM profiles due to numerical artifacts \citep{shea13} that may be
misinterpreted as core formation by stellar feedback in low-resolution
hydrodynamical simulations.

If the addition of baryons is unable to fully address the small scale issues of CDM, 
it may be that there is actually no problem but rather it is an illusion caused by observational effects (as suggested by \citet{pineda17} and references therein); another 
approach is to 
consider different dark matter properties.
Some alternative DM models are self-interacting dark matter (SIDM)
\citep{spe00}, ultra-light (scalar field/Bose-Einstein Condensate) dark matter
\citep{sin94,lee96,guz00,mat11,rob13,sua14,mocz17}, and warm dark matter models
\citep{lov14,mac12}. In this work, we focus on the SIDM model and consider the
simplest option: identical dark-matter particles undergoing isotropic,
velocity-independent, elastic, hard-sphere scattering with a cross section of
$\sigma$.  The scattering rate per
particle scales as $\Gamma(r) \sim \rho(r) (\sigma/m)v_{\mathrm{rms}}$,
depending on the local mass density $\rho$ and the r.m.s. speed of DM particles
$v_{\mathrm{rms}}$.  Current constraints from DMO simulations of dwarf haloes
that include self-interactions suggest that
$0.5\, \cmg<\sigma/m<5\, \cmg$
can lead to cores of \textit{O}($1\rm \, kpc$) in their centers
\citep{elb15,fry15} and thereby 
alleviating the CDM problems without the need of the baryonic component in dark
matter dominated systems.  

While the effects of baryons on CDM haloes and the effects of self-interactions
in DMO simulations have been examined extensively in the context of CDM's
small-scale ``crisis'', much less work has explored the effects of baryonic
physics and self-interactions simultaneously. \citet{vog14} and \citet{fry15}
both found that galaxies with $\mstar(z=0) \approx 10^8\,\msun$ simulated in
SIDM with full hydrodynamics resulted in galaxies that were not appreciably
different from CDM hydrodynamic (CDM-Hydro) simulations. However, it is not
obvious that this is true at all stellar masses, as $\mstar \sim 10^8\,\msun$ is
near the peak of the core formation efficiency in CDM-Hydro simulations
\citep{tollet16,chan15,dicintio14}. It is especially interesting to consider
systems with $\mstar \sim 10^5-10^6\,\msun$, as most theoretical work indicates
such galaxies should retain their NFW cusps even when incorporating baryonic
feedback (\citet{fitts16} and references therein).

In this paper, we address the robustness of SIDM predictions using simulations
of a sample of 4 low-mass dwarf galaxies that incorporate realistic galaxy
formation and stellar feedback models. The simulations are discussed in
Section~\ref{sec:sims}. In Section~\ref{sec:results}, we compare the results for
the SIDM-DMO, CDM-DMO, and their corresponding hydrodynamical versions. Section
~\ref{sec:conclusion} presents our main conclusions. We adopt a cosmological
model with parameters $\sigma_8=0.801$, $\Omega_\Lambda=0.734$,
$\Omega_m=0.266$, $\Omega_b=0.0449$, $n_s=0.963$, and $h=0.71$ \citep{komatsu11}
throughout this work.

\section{Simulations}
\label{sec:sims}
The starting point for our investigation is the cosmological hydrodynamical
zoom-in simulations of \citet{fitts16}.  The Fitts et al. suite comprises 15
isolated haloes, with $\mhalo(z=0)\approx 10^{10}\,\msun$ and a diversity of
assembly histories and $z=0$ concentrations, chosen from periodic parent volumes
with box sizes of 35 Mpc each. The simulations were all run as part of the {\tt
  FIRE}\footnote{http://fire.northwestern.edu} project \citep{hop14} and adopt
the {\tt FIRE}-2 model \citep{hop17}.  Accordingly, all of the simulations were
performed with the
GIZMO\footnote{http://www.tapir.caltech.edu/$\sim$phopkins/Site/GIZMO.html}
code, and hydrodynamical versions use the mesh-free finite-mass (MFM) method in
GIZMO.  The high-resolution simulations have fixed gravitational
softenings\footnote{We use the Plummer equivalent softening, the real region
  that is softened is $2.8\, \epsilon$.} of $\epsilon_{\mathrm{dm}}=35\,\pc$ for
the DMO, and $\epsilon_{\mathrm{dm}}=35\,\pc$ and $\epsilon_{\star}= 3\,\pc$
physical for the stars.  The gas smoothing is fully adaptive and is the same for
the hydrodynamic kernel and the gravitational softening; the minimum physical
softening is $h_{\mathrm{gas}}=1.4\,\pc$. The DM particle mass is
$m_{\mathrm{dm}} \approx 3000\,\msun$ for DMO and
$m_{\mathrm{dm}}\approx 2500\,\msun$ for Hydro runs;
$m_{\mathrm{gas,initial}}\approx 500\,\msun$ and the initial stellar mass is
similar to the gas mass.  At $z=0$, the haloes host galaxies with
$5.6<\log_{10}(\mstar/\msun) < 7.1$.

From this suite, we have selected four haloes that span the full range of $z=0$
stellar masses.  Following the naming convention in \citet{fitts16}, we
resimulated haloes m10b, m10d, m10f, and m10k with a self-interaction cross
section of $\sigma/m=1 \,\cmg$ using the SIDM implementation of \citet{rocha13}.
This method considers interactions between pairs of phase-space patches,
  taking into account the collision term in the Boltzmann equation.  For each
halo simulated in SIDM, we perform a DMO version and a version with full FIRE-2
galaxy formation physics. Note that we therefore have 4 versions of each of the
4 haloes: CDM-DMO, CDM-Hydro, SIDM-DMO, and SIDM-Hydro. The implementation of
galaxy formation physics is identical for all hydrodynamic runs (CDM and SIDM).
Our high-resolution runs do not suffer from numerical relaxation for radii
larger than $200 \,\pc$ based on to the convergence criterion of \cite{power03};
we adopt this value as our convergence radius in the density profiles.  To
identify the haloes in the simulations, we use the public code ROCKSTAR
\citep{ber13}. We tested the robustness of the ROCKSTAR-determined centers using
the Amiga Halo Finder code \citep{kno09} and found no distinguishable
differences in the converged region; for SIDM haloes, we found ROCKSTAR centers
to be more accurate.

\begin{figure}
\resizebox{250pt}{225pt}{\includegraphics{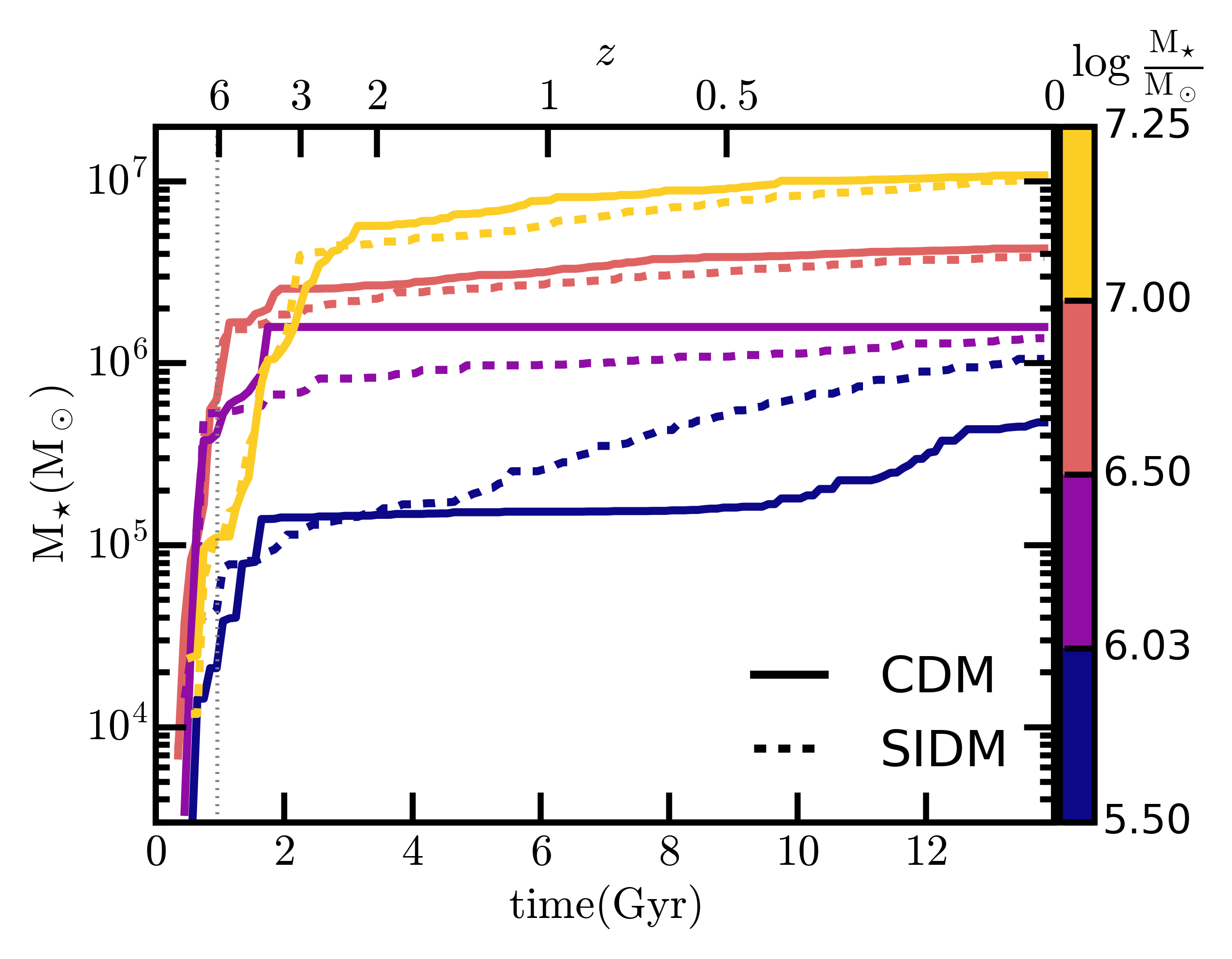}}
\caption{Evolution of the cumulative star formation history for each of our
  simulated galaxies. The four CDM galaxies (solid lines) are from the sample of
  \citet{fitts16}, corresponding to: m10b (blue), m10d (purple), m10f (red), and
  m10f (yellow). Dashed-lines represent our SIDM simulations with the same
  initial conditions as the CDM haloes. Galaxies are colored according to their
  stellar mass at redshift $z=0$ (see Table~\ref{tab:table1} for the exact
  values). We use the same color code in every figure. The dotted vertical line
  is where reionization ends in the simulations.}
\label{fig:figure1}
\end{figure}

\begin{figure}
\resizebox{!}{210pt}{\includegraphics{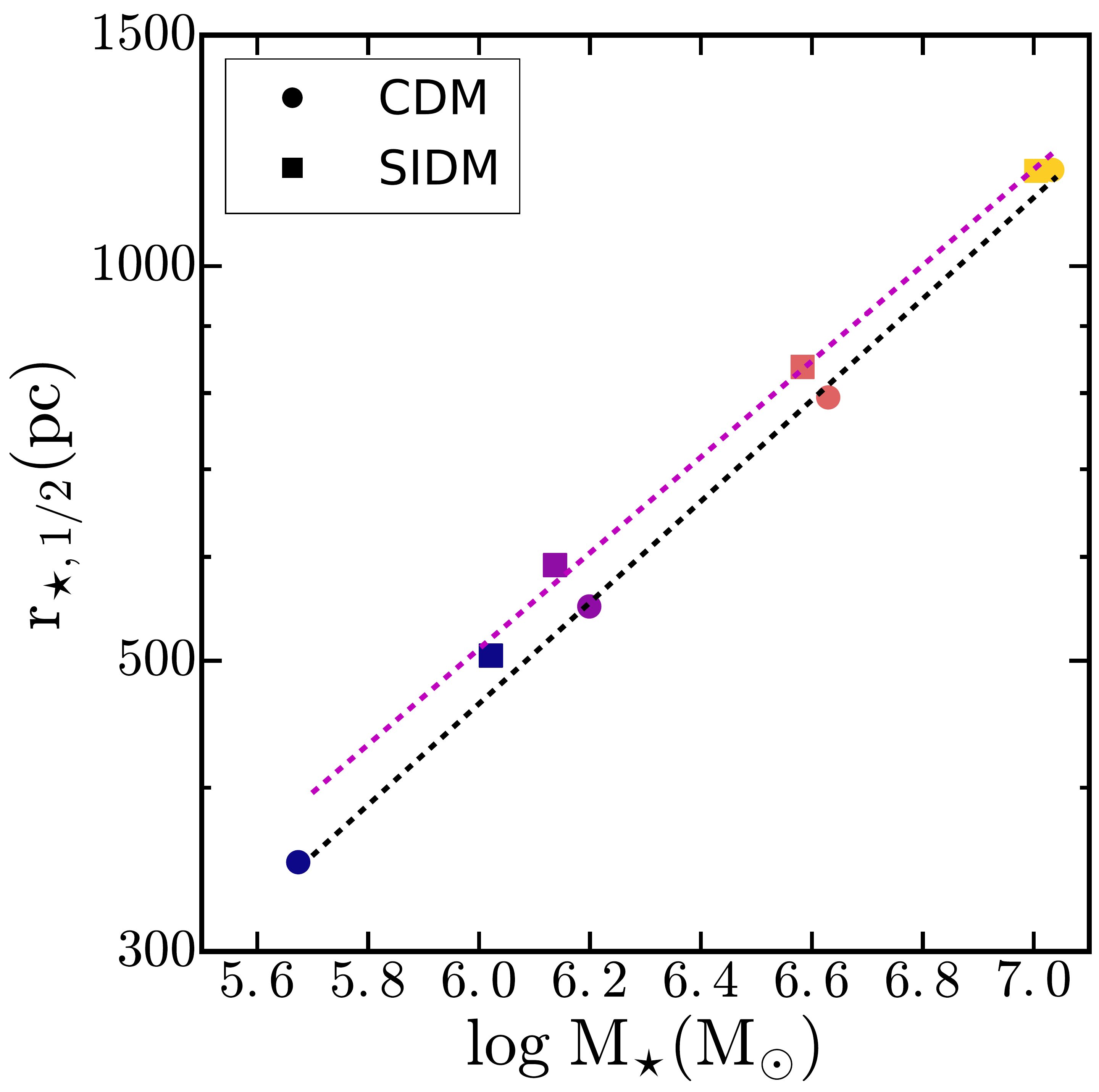}} 
\caption{Effective 3D stellar half-mass radius vs total $\mstar$ within $0.1\,  r_{\rm{vir}}$ for the CDM (circles) and SIDM
  (squares) simulations. Note that the relationship for CDM galaxies and SIDM galaxies is very similar, with SIDM galaxies slightly larger at fixed stellar mass.  
  The CDM galaxies are well fit 
  by (dashed black line) $r^{\rm {cdm}}_{\star,1/2} \propto (M^\mathrm{cdm}_{\star})^{0.386}$ while the SIDM
galaxies follow a similar relation (dashed magenta line) $r^{\rm {sidm}}_{\star,1/2}  \propto (M^\mathrm{sidm}_{\star})^{0.365}$.}  
\label{fig:figure2}
\end{figure}

\begin{table*}
	\caption{Properties at $z=0$ for the simulated galaxies in CDM (values taken from \citet{fitts16}) and in SIDM. Columns: (1) Halo name used in the suite of \citet{fitts16}; (2) Halo virial mass; (3) Maximum amplitude of the circular velocity; (4) Galaxy stellar mass (defined as $M_{\star}(<0.1 \, R_{\rm vir}$)); (5) 3D stellar half-mass radius; (6) Maximum of the circular velocity (DMO, after correction for cosmic baryon fraction $f_{\rm b}$); (7) Ratio of virial mass in hydro run to the virial mass in DMO run (DMO virial mass corrected for $f_{\rm b}$).}
	\label{tab:table1}
	\resizebox{.9\textwidth}{!}{
	\begin{tabular}{lccccccr} 
		\hline
		 & \multicolumn{6}{c}{CDM}  \\
		\hline
		Halo  & $M_{\rm vir}$ & $V_{\rm max}$ & $\mstar$ & $r_{\star,1/2}$ & $V^{\rm DMO}_{\rm max}$ & $M_{\rm hydro}/M_{\rm dmo}$ \\ 
		  & $\msun$ & $\rm{km\,s^{-1}}$ & $\msun$ & $\pc$ & $\rm{km\,s^{-1}}$ & --\\ 
		\hline
		m10b & $9.29 \times 10^9 $ & $31.5$ & $4.65 \times 10^5 $ &  $340$ & $34.8$ & 0.96 \\ 
		m10d & $8.43 \times 10^9 $ & $32.1$ & $1.53 \times 10^6$  & $530$ & $37.6$ & $0.98$ \\
		m10f &  $8.56 \times 10^9 $ & $35.7$ & $4.11 \times 10^6$  & $750$ & $41.2$ & $0.94$ \\ 
		m10k & $1.15 \times 10^{10}$ & $38.2$ & $1.04 \times 10^7$ & $1140$ & $43.5$ & $0.96$ \\
		\hline
		 & \multicolumn{6}{c}{SIDM} \\
		\hline
		m10b & $8.13\times 10^9$ & $30.8$ & $1.05 \times 10^6$ &$504$ & $31.8$ & $0.90$ \\ 
		m10d & $8.10\times 10^9$  & $33.1$ & $1.37 \times 10^6$  & $591$ & $34.5$ & $0.94$ \\ 
		m10f &  $8.39\times 10^9$  & $35.7$ & $3.83\times 10^6$  & $838$ & $38.8$ & $0.93$ \\ 
		m10k & $1.12\times 10^{10}$  & $37.6$ & $1.01\times 10^7$ & $1260$ & $40.4$ & $0.94$ \\ 
		\hline
	\end{tabular}
	}
\end{table*}

\begin{figure*}
\begin{tabular}{ll}
\resizebox{245pt}{!}{\includegraphics{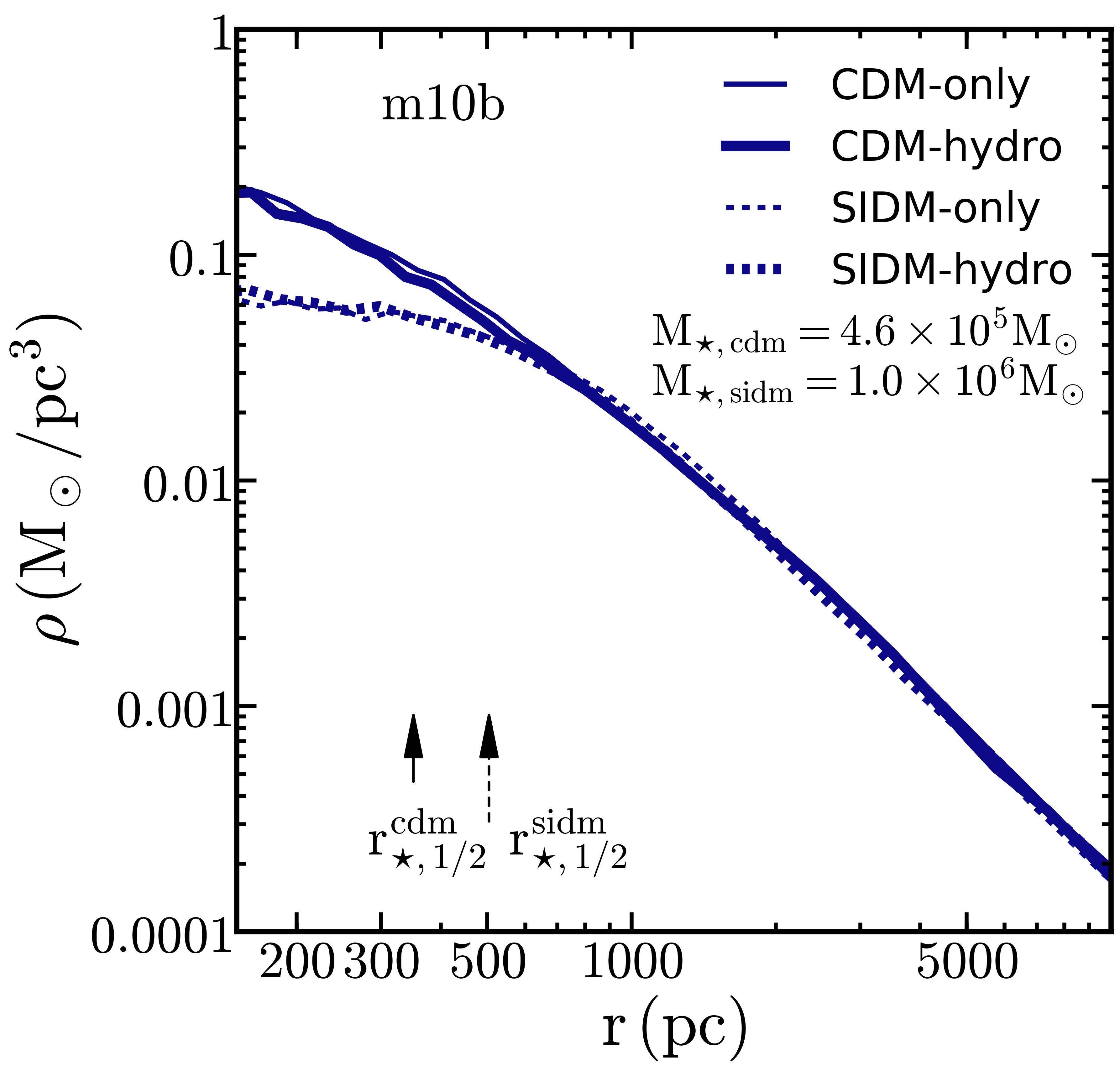}} &
\resizebox{245pt}{!}{\includegraphics{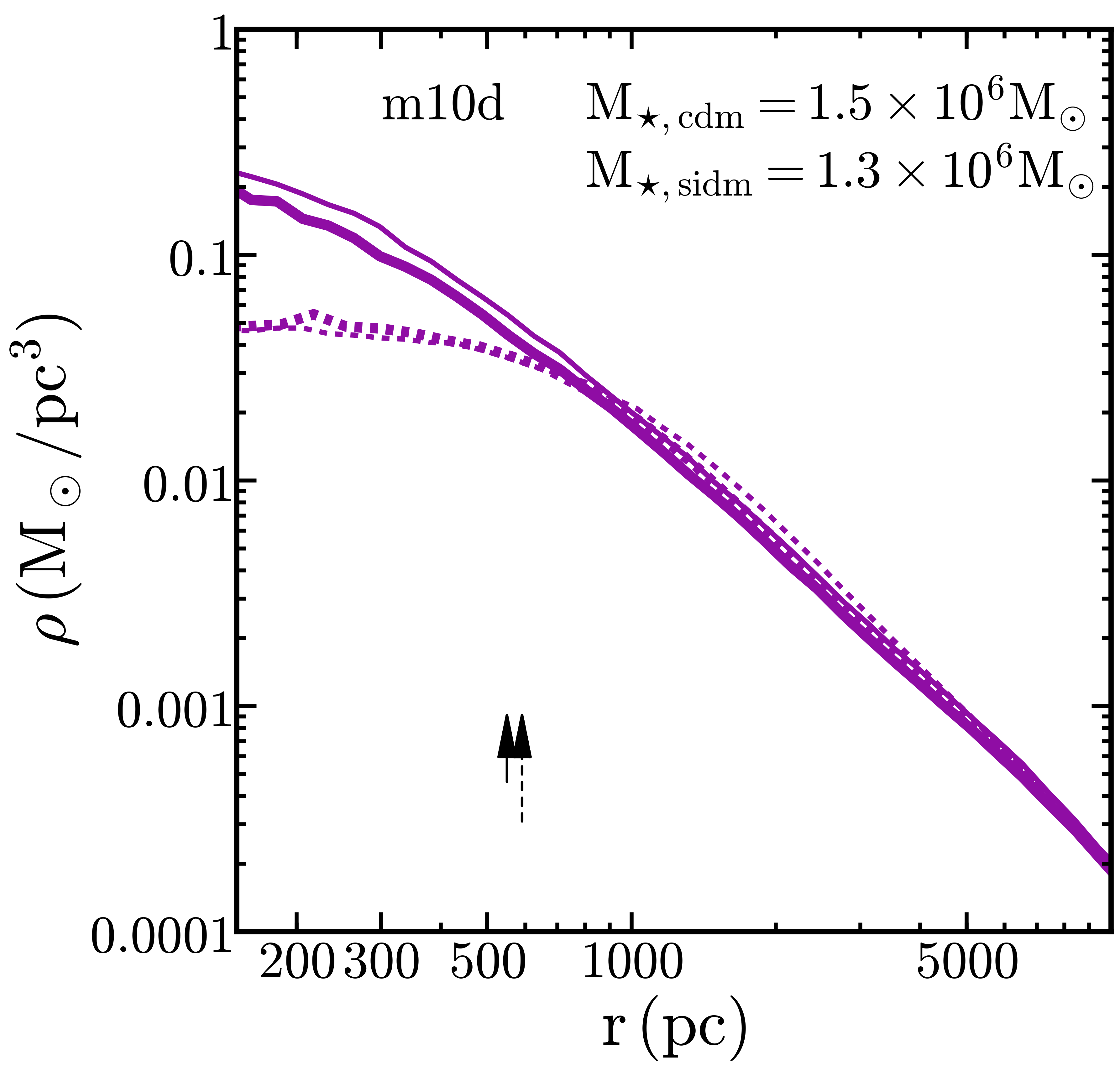}} \\
\resizebox{245pt}{!}{\includegraphics{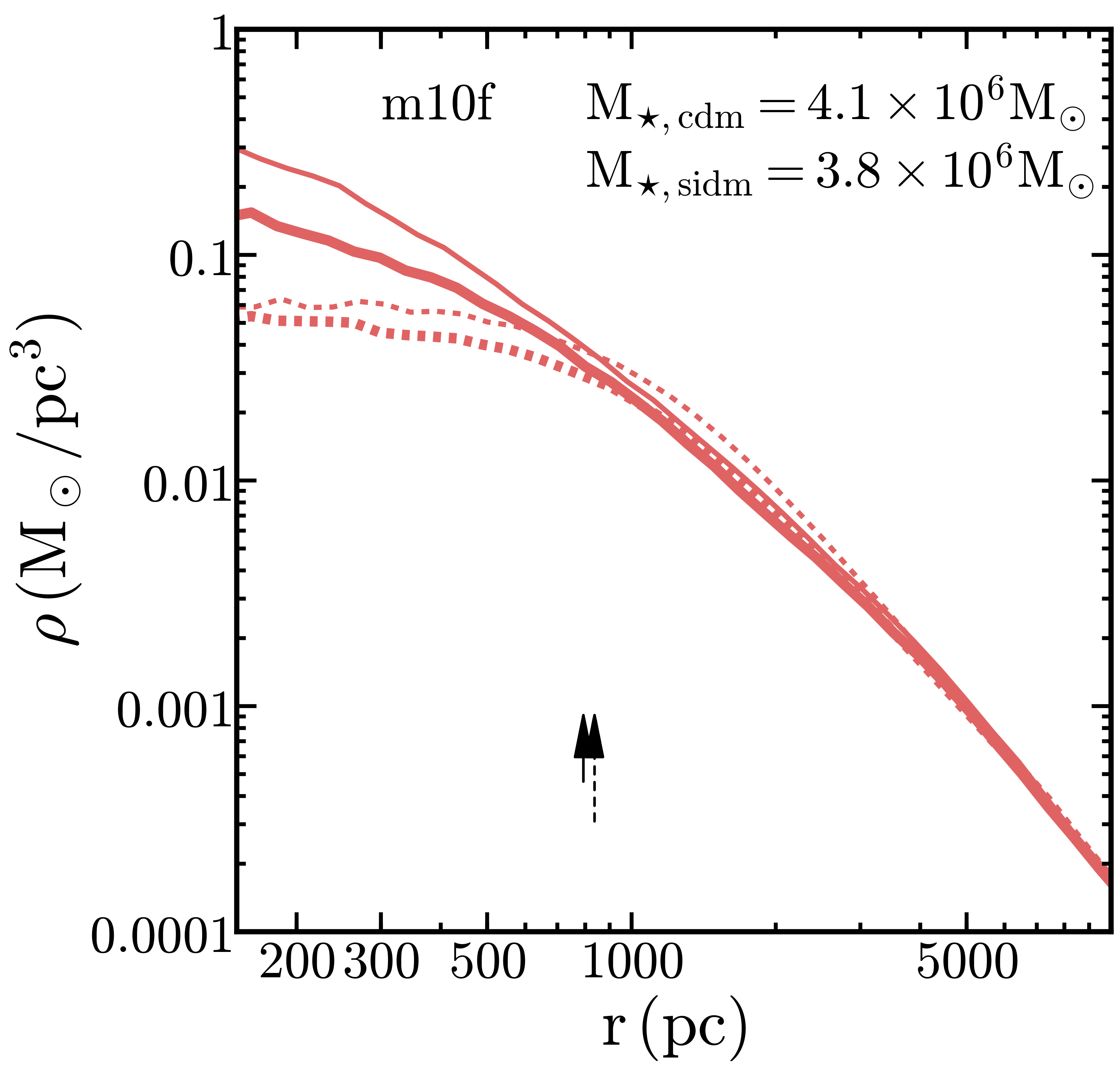}} &
\resizebox{245pt}{!}{\includegraphics{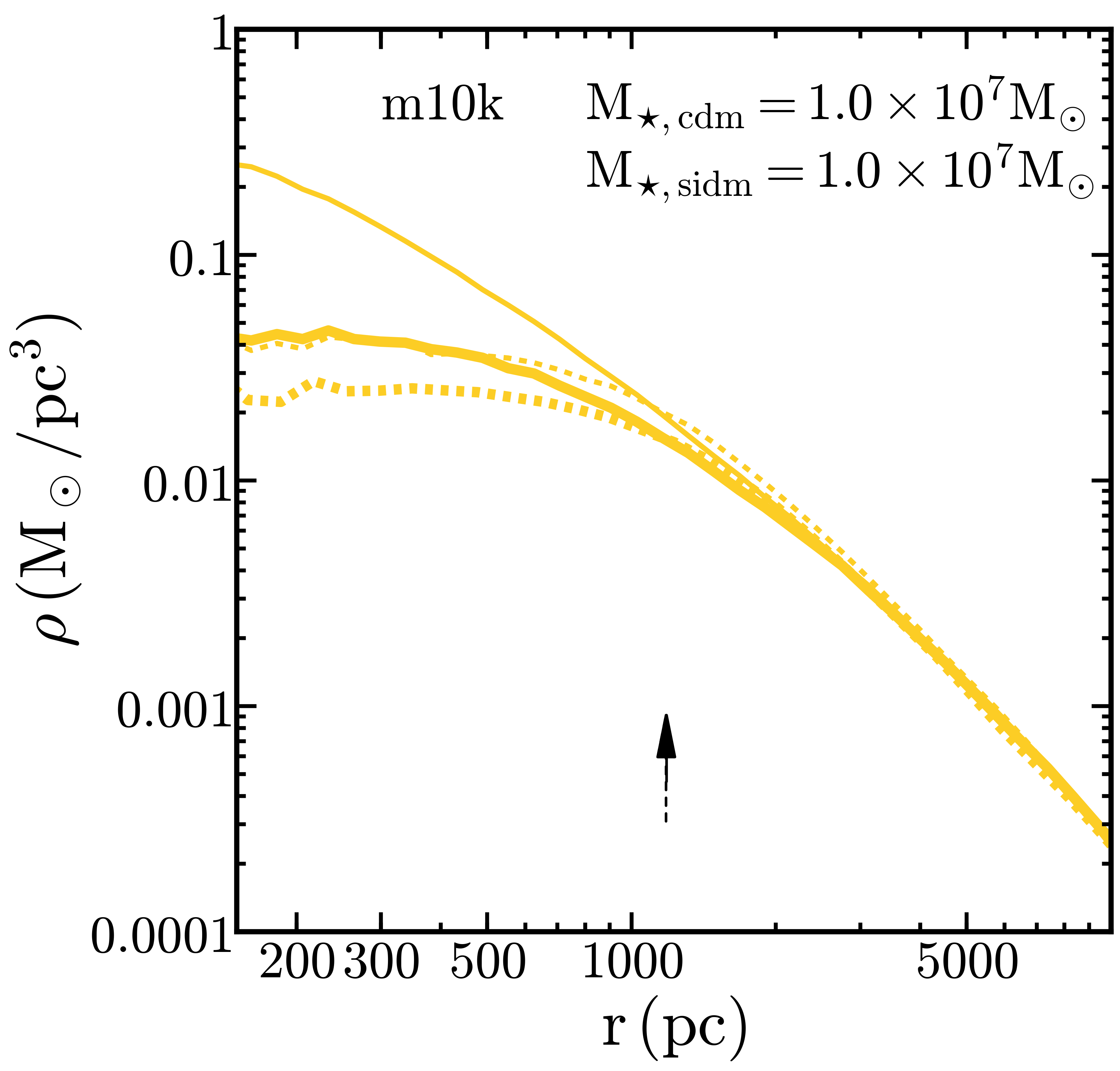}} \\
\end{tabular}
  \caption{Dark matter density profiles for the four SIDM (dashed lines) and CDM (solid lines) simulations. Profiles of the
hydrodynamical simulations with the FIRE physics are shown with thick lines and dark matter-only (DMO) simulations 
are shown with thin lines. Also shown with arrows in each panel are the
effective stellar mass radii ($r_{\star,1/2}$). The colors are the same as in Fig.1; each panel is labeled with the name of the halo and its stellar mass at $z=0$ in both dark matter models (see Table~\ref{tab:table1} for a summary of the individual properties.)}
\label{fig:figure3}
\end{figure*}

\begin{figure*}
\begin{tabular}{cc}
\resizebox{240pt}{!}{\includegraphics{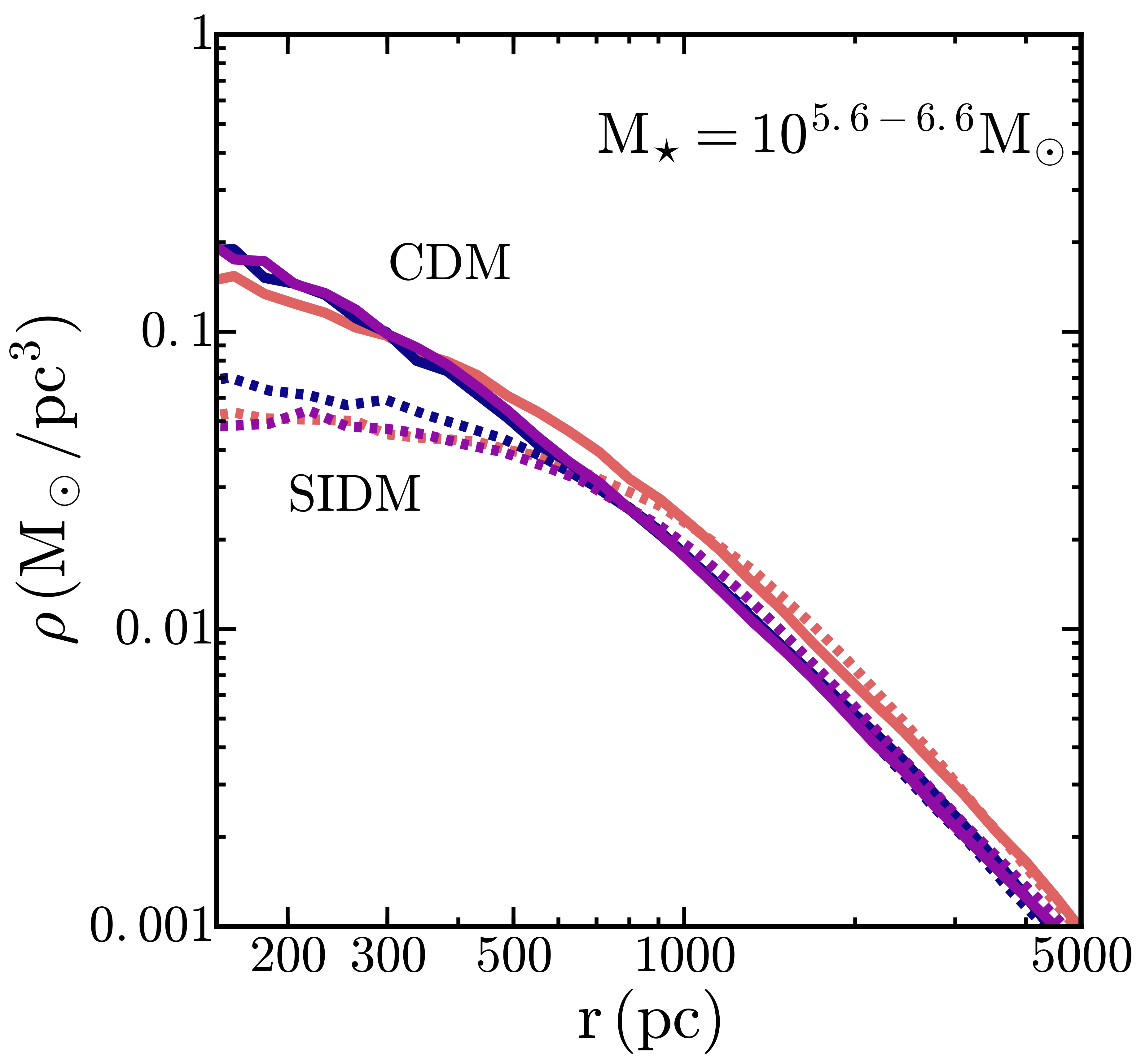}} &
\resizebox{240pt}{!}{\includegraphics{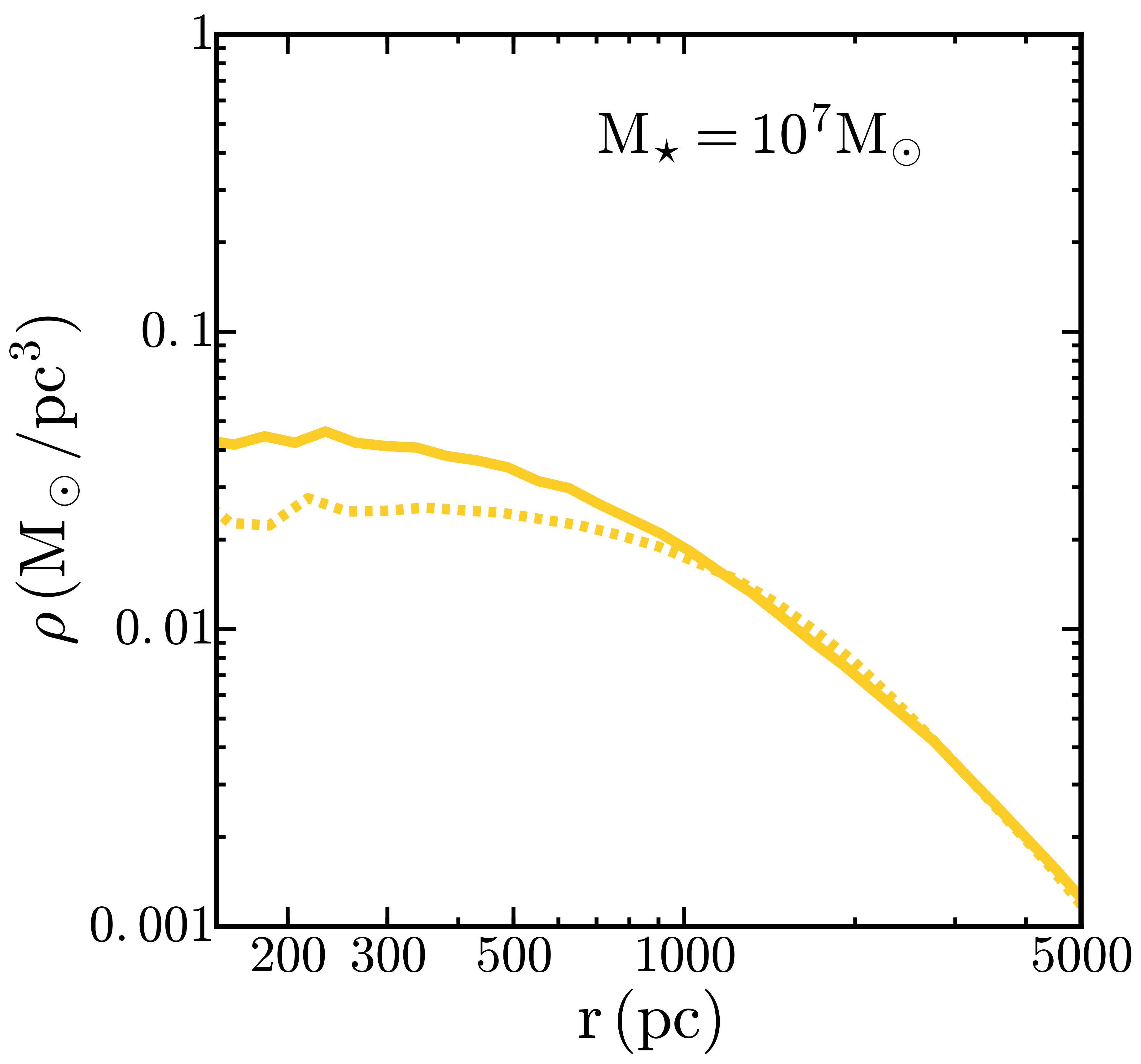}} \\
\end{tabular}
  \caption{Dark matter density profiles for our FIRE-2 hydro simulations that form $\mstar=10^{5.6-6.6}\msun$ (left panel) and for the most massive galaxy ($M_\star = 10^{7} \msun$; right panel). 
Dwarf galaxy haloes in CDM retain their cusp for $\mstar<10^{6.6}\msun$; only in our most massive galaxy both CDM and SIDM display a large core ($\sim 1$ kpc).}
\label{fig:figure4}
\end{figure*}

\begin{figure*}
\begin{tabular}{ll}
\resizebox{245pt}{!}{\includegraphics{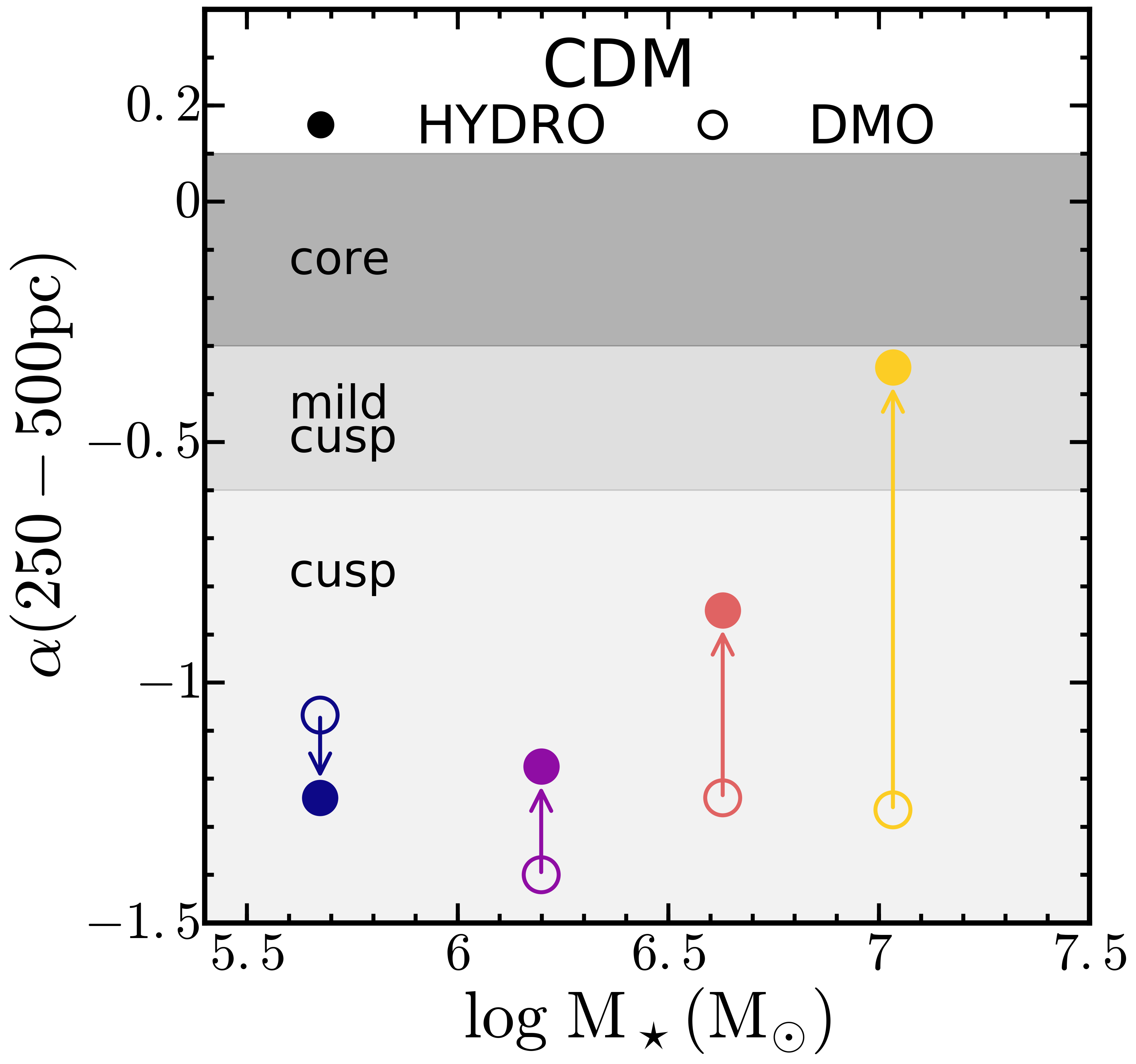}} &
 \resizebox{245pt}{!}{\includegraphics{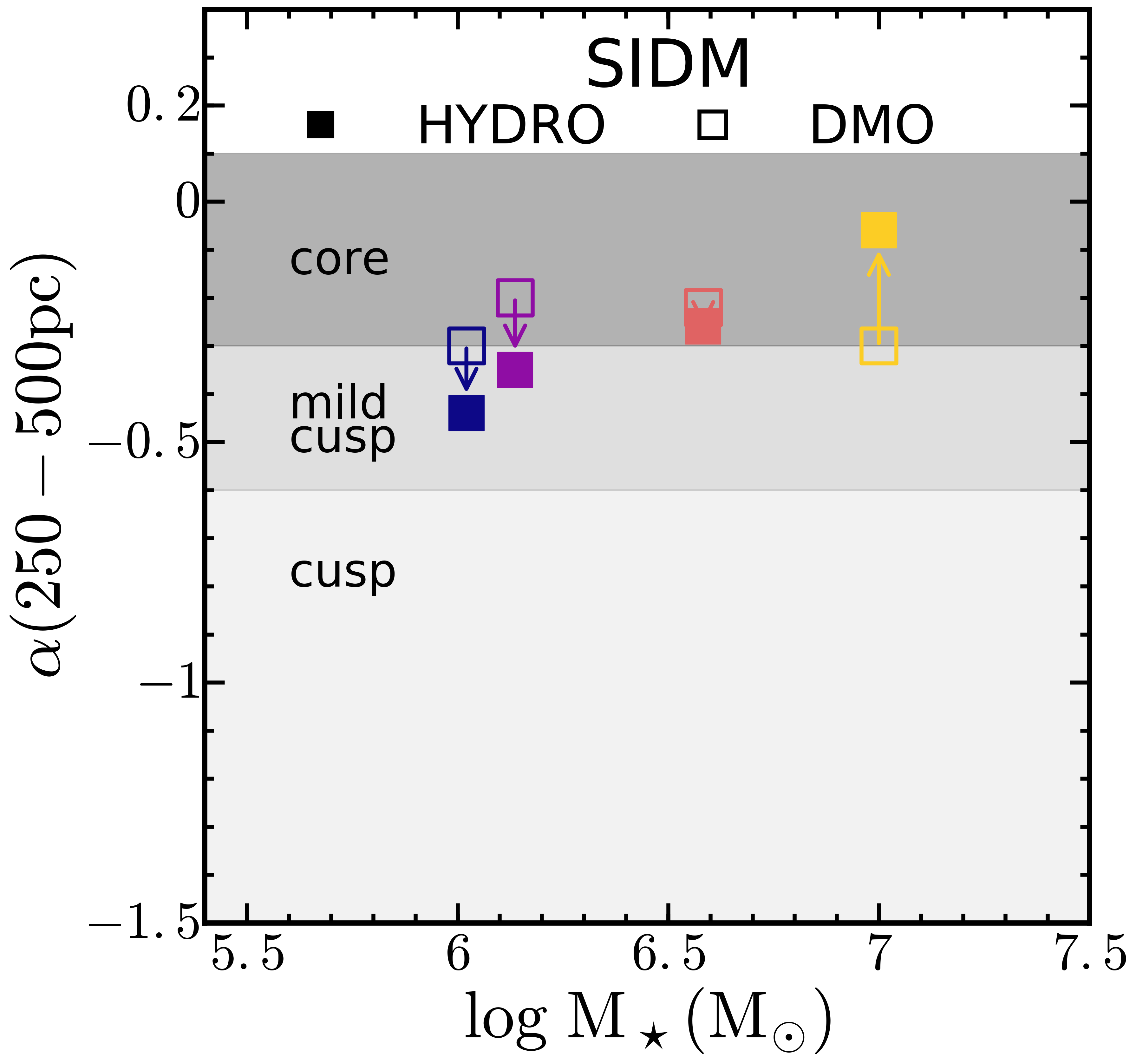}} \\
 \resizebox{245pt}{!}{\includegraphics{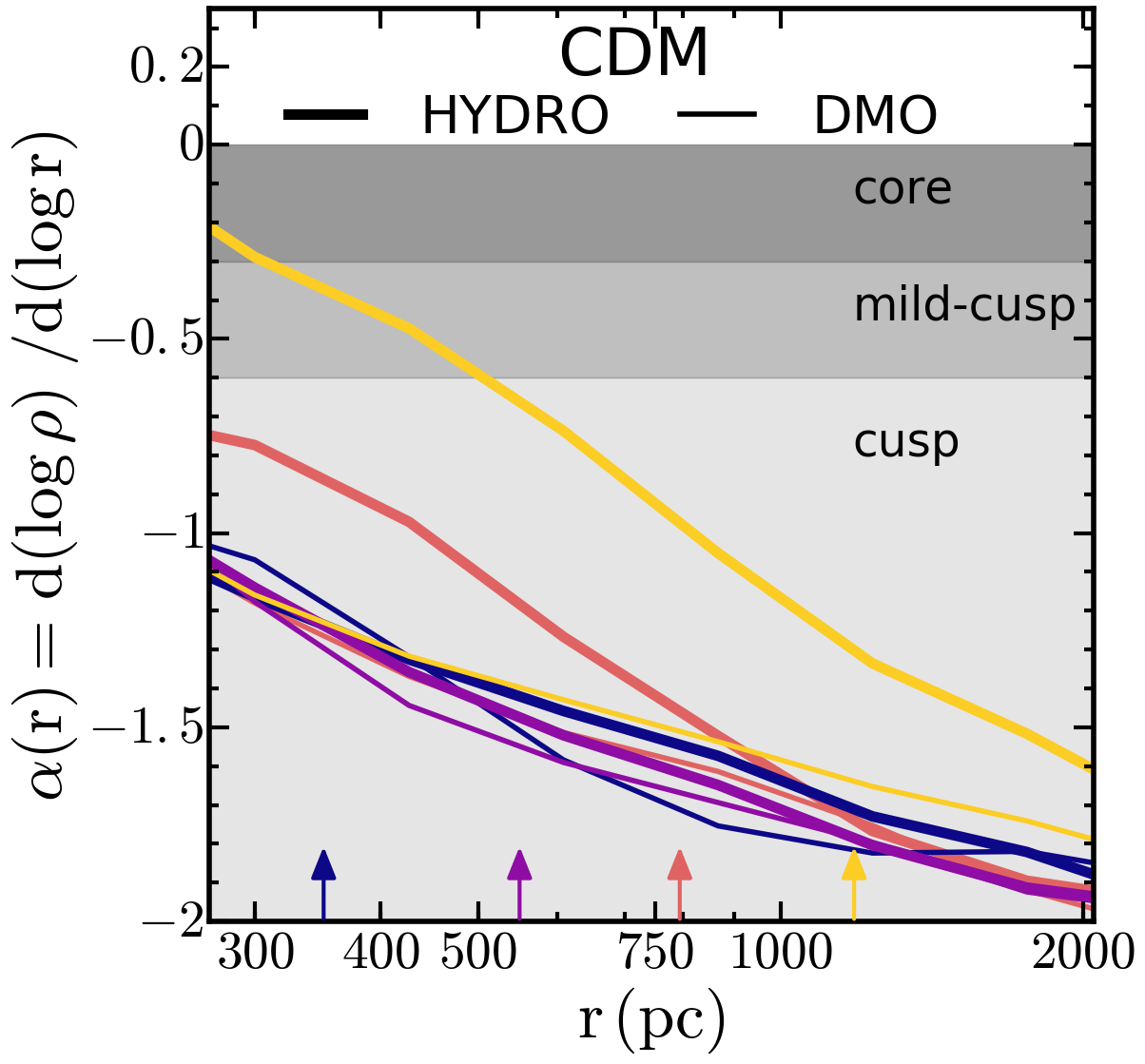}} &
\resizebox{245pt}{!}{\includegraphics{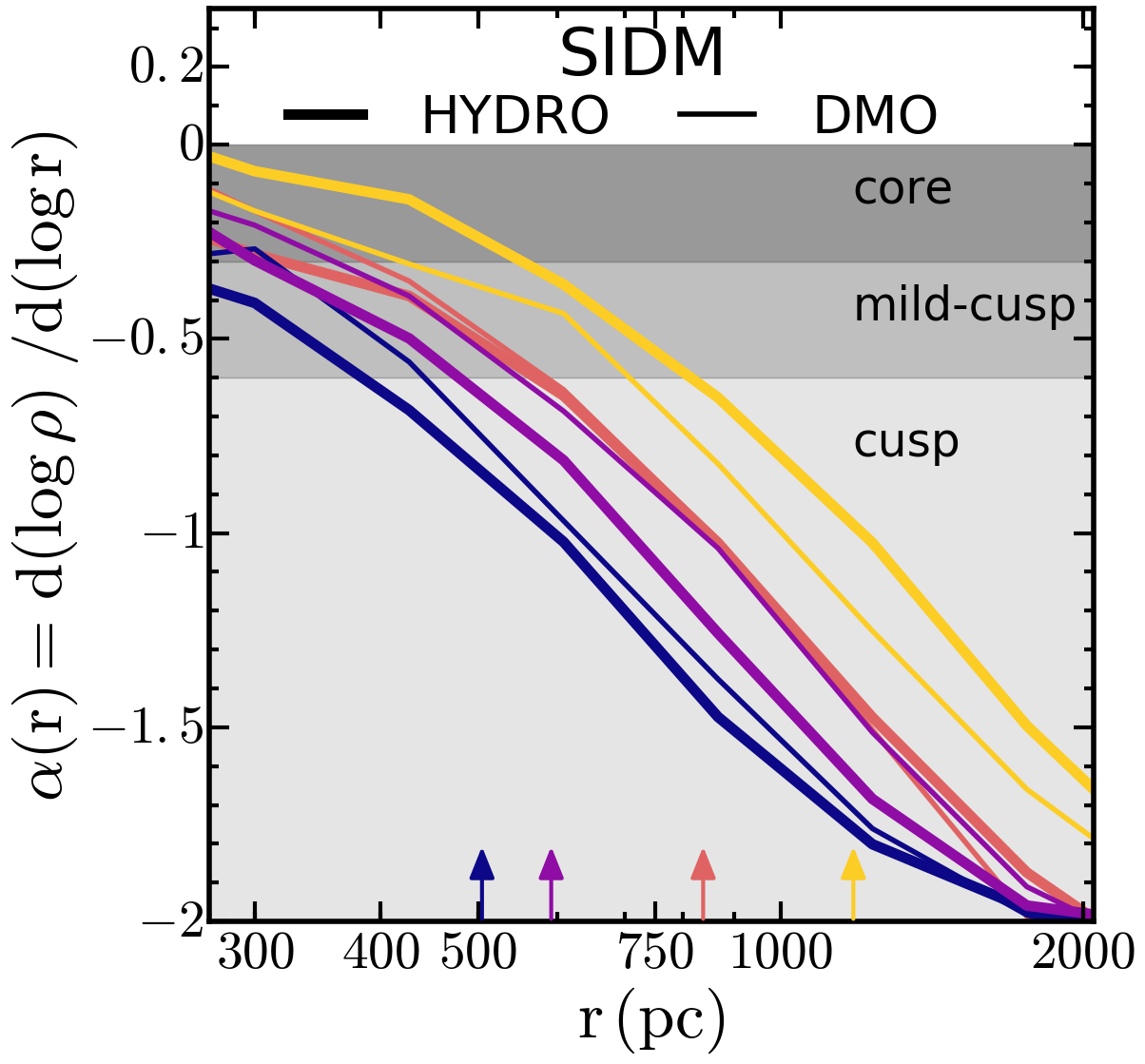}} \\
\end{tabular}
  \caption{Upper row: slopes of DM density $\alpha$ as a function of halo mass for the simulated CDM haloes (left column) 
and SIDM haloes (right column). The slopes were obtained by fitting the DM density profiles in the range $250-500 \, \pc$ ($0.5-1\% r_{\mathrm{vir}}$). 
Filled symbols correspond to simulated haloes with FIRE and empty symbols are with DMO, for the latter we used the same stellar 
mass (and color) as their hydro simulation for an easier comparison of the slopes. 
Lower row: slope of DM density vs radius for CDM (left column) and SIDM (right column) haloes, thick lines represent the simulations with FIRE and thin lines the DMO ones (we use the same color for the respective DMO run). The arrows at the bottom mark the effective half-mass radius, $r_{\star,1/2}$, for its associated hydro simulation (identified by the same color of the arrow). 
The horizontal dark gray 
region is where the density profile is flat enough so that we call it a core ($-0.3 \leq \alpha< 0.1$), below (light gray) is the mild-cusp 
region ($-0.6 \leq \alpha<-0.3$) and at the bottom is the cusp region ($\alpha<-0.6$).}
\label{fig:figure6}
\end{figure*}

\begin{figure*}
\centering
\begin{tabular}{ll}
\resizebox{238pt}{!}{\includegraphics{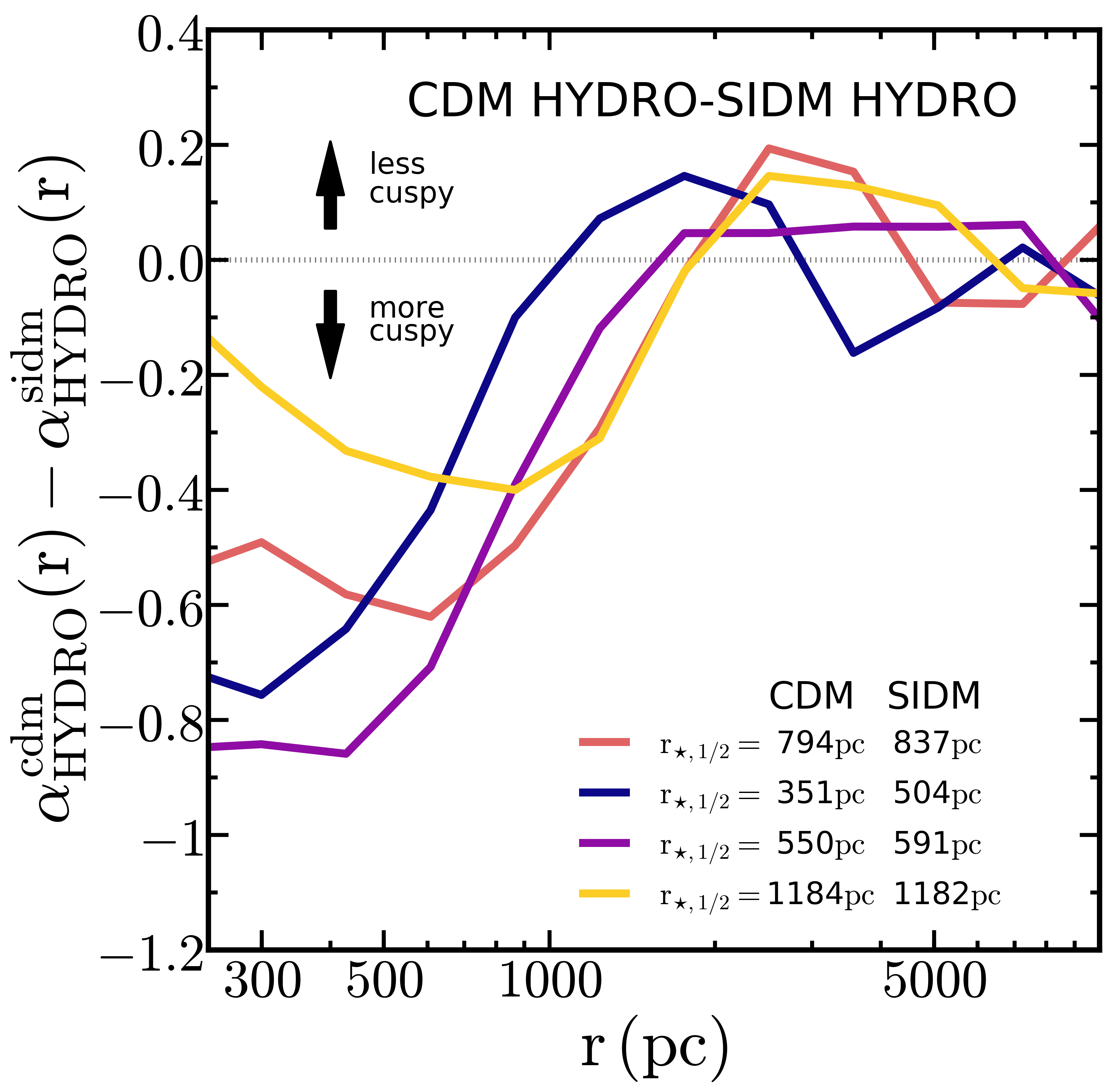}} &
 \resizebox{232pt}{229pt}{\includegraphics{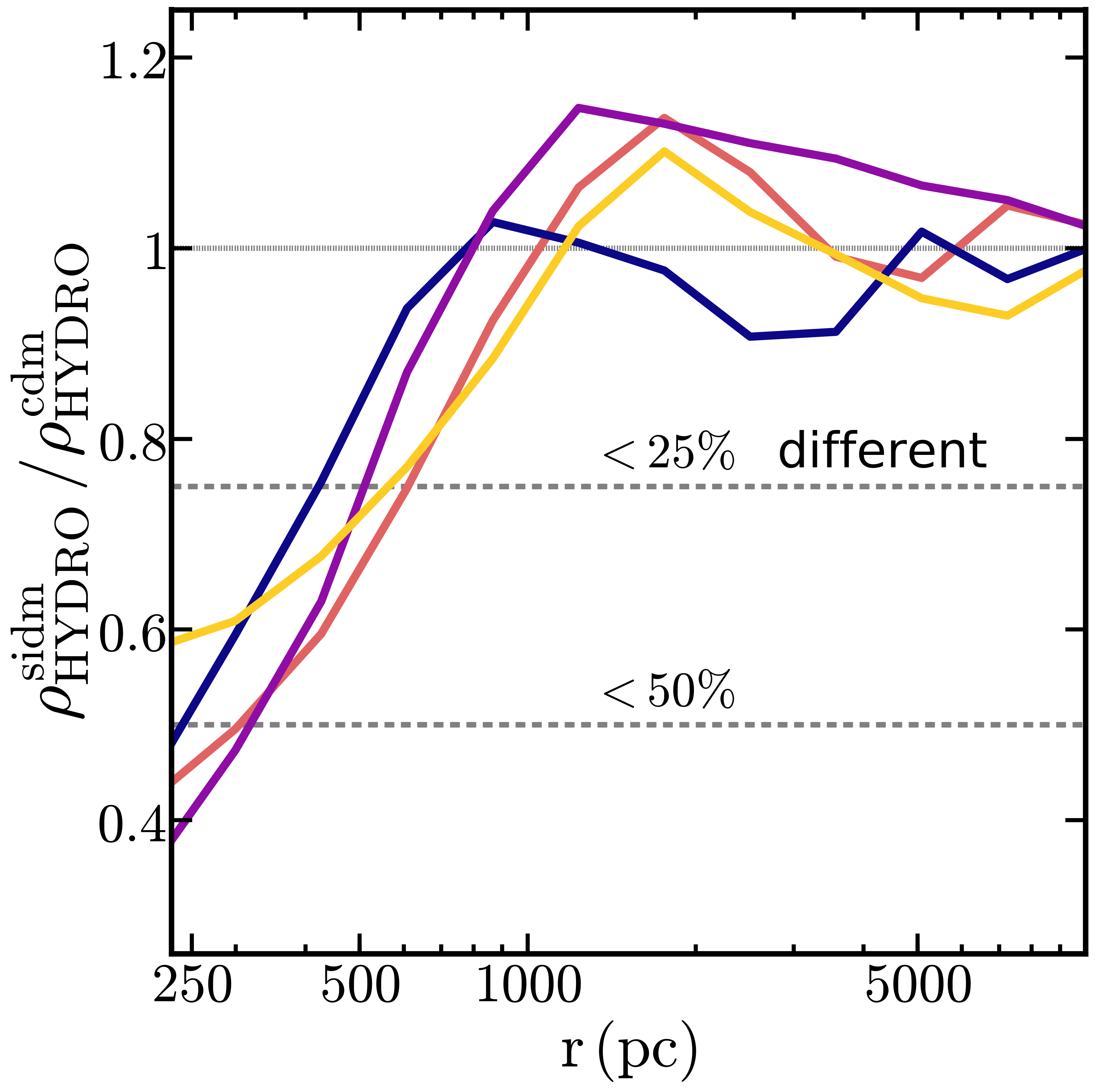}} \\
\end{tabular}
 \caption{Left: Relative change of the DM density slope between the CDM and SIDM-Hydro simulations vs radius. The largest difference appears in the most massive galaxy starting at the effective stellar radius ($r_{\star,1/2}$), this is also seen for the other galaxies at their respective $r_{\star,1/2}$, as can be inferred from their values in the labels. Right:
Dark matter density ratio of the SIDM-Hydro simulation and its CDM-Hydro pair. The gray dashed lines show the transition limit  above which the ratio of the density profiles differs in less than $<25\%$ and $<50\%$, respectively. SIDM galaxies become less dense towards the center reaching a $\geq 25\%$ difference from their CDM counterparts at $\sim 500 \, \pc$. 
} 
\label{fig:figure7}
\end{figure*}

\section{Results}
\label{sec:results}
\subsection{Global Properties}
\label{subsec:global}

Figure~\ref{fig:figure1} shows the star formation histories of the four galaxies
simulated in this work, with SIDM runs shown as dashed lines and their CDM
counterparts shown as solid lines.  These galaxies span the range of star
formation histories in the \citet{fitts16} sample, which show a variety similar
to those observed in nearby dwarf galaxies \citep{skill14,cole14}. The three
most massive galaxies in CDM have very similar star formation histories in SIDM
(and very similar final $\mstar$). Only the lowest-mass halo exhibits a notable
difference: while the CDM-Hydro simulation shows an extended pause in star
formation from $\sim 2$ Gyr to $\sim 9$ Gyr of cosmic time, the SIDM-Hydro
simulation forms stars continuously and ends up with twice as many stars at
$z=0$. It is not clear why these two galaxies show the most significant
differences but it may be related to the enhanced sensitivity of star formation
in small dwarfs that are most susceptible to UV background feedback
\citep{benitez17}; a detailed investigation of different types of feedback and
their effects on galaxy formation in various dark matter models will be
presented in future work.

The $z=0$ properties of our haloes and galaxies are summarized in
Table~\ref{tab:table1}.  Properties listed include the halo virial
masses\footnote{We define all virial quantities using the \citet{bryan98}
  definition of the virial overdensity. For our chosen cosmology
  $\Delta_{\rm vir}=96.45$ (relative to $\rho_{\rm{crit}}$ at $z=0$.)} at $z=0$ in
each baryonic run, maximum circular velocities for both DMO and hydro runs, and
the ratio of the virial mass in the hydrodynamic runs to those in the DMO
runs. The ratio $M_{\rm hydro}/M_{\rm dmo}$ is defined such that DMO virial mass
assumes a loss of all baryonic matter: $M_{\rm dmo} = (1-f_b) M_{\rm vir}$.  We
note that the quantities are generally fairly stable between the CDM and SIDM
runs.

Table~\ref{tab:table1} also lists the 3D stellar half-mass radius,
$r_{\star, 1/2}$, for each galaxy.  The relationship between stellar mass and
galaxy $r_{\star, 1/2}$ is plotted in Figure~\ref{fig:figure2}. Results for CDM
are shown as circles, while results for SIDM simulations are plotted as
squares. Note that both dark matter models produce a similar stellar mass
vs. galaxy size relationship.  Even the lowest-mass halo, which forms twice as
many stars in SIDM than in CDM, also falls on the stellar mass-size relation in
agreement with the rest of the simulations.  In fact, all of the simulations,
both CDM-Hydro and SIDM-Hydro, lie on a $\mstar-r_{\star,1/2}$ relationship that
is well approximated by
$r_{\star, 1/2} \approx 456\, {\rm pc} \left(M_\star/10^6 \msun \right)^{0.37}$.
Future work using different SIDM cross sections will reveal whether this
similarity predicted by our galaxy size relation in SIDM and CDM holds beyond
the specific cross section adopted here.

\subsection{Density profiles}
\label{subsec:rho}
In Figure~\ref{fig:figure3}, we show the DM density profiles for all of our
simulations. Each panel shows the DMO profiles\footnote{DMO density profiles are
  corrected for the cosmic baryon fraction as in \citet{fitts16}.} (thin lines)
and hydro profiles (thick lines). In both cases, we plot results for CDM (solid)
and SIDM (dashed) simulations. The arrow in each figure indicates the stellar
half-mass radius ($r_{\star,1/2})$ of the host galaxy. The galaxy that forms the
lowest total stellar mass is shown in the upper left, while the most massive
galaxy is shown in the bottom right panel. Fig.~\ref{fig:figure3} shows that, in
all cases, DMO simulations exhibit central density \textit{cusps} in CDM (thin
solid lines) and central density \textit{cores} in SIDM (thin dotted lines).

Although all 4 haloes have nearly the same virial mass at $z=0$, they have
somewhat different assembly histories, leading to different concentrations and
values of $V_{\rm max}$ \citep{fitts16}. These differences are further
reflected in the core sizes seen in the SIDM-DMO runs in
Fig.~\ref{fig:figure3}. The latest-forming, lowest-concentration haloes have
lower central densities in CDM; lower central densities result in fewer DM
interactions, as the interaction rate $\Gamma$ scales as
$\Gamma \propto \rho\,(\sigma/m)\,v$. Thus, the smaller $V_{\rm max}$ haloes end
up with smaller SIDM-induced cores in the DMO runs.

As argued in \citet{fitts16}, the more centrally-concentrated haloes are also
the ones that form more stars in the CDM-Hydro runs, as they can accumulate more
gas earlier and their central gravitational potentials are deeper, helping to
offset the effects of later reionization feedback. In the CDM-Hydro runs,
increasing the stellar mass also enhances core formation via star formation
feedback. The density profile in the lowest-mass galaxy in the suite has no
discernible difference when including hydrodynamics in CDM (upper left in
Fig.~\ref{fig:figure3}); effects are still very small at
$\mstar \sim 10^6\,\msun$ (upper right) but are beginning to become apparent
when $\mstar \sim 4\times 10^6\,\msun$ (lower left). The most massive galaxy
(lower right), with $\mstar \sim 10^7\,\msun$, has a pronounced density core in
the CDM-Hydro run.

When including the effects of both galaxy formation and self-interactions, the
situation changes both qualitatively and quantitatively. In all cases, the
difference in density structure between SIDM-DMO and SIDM-Hydro simulations are
relatively small, and the effects are smaller than in the equivalent CDM-Hydro
runs in every case. The largest effects for SIDM-Hydro are seen in the most
massive galaxy, where the core density is reduced by $\sim 40\%$ relative to the
SIDM-DMO run (the core radius remains the same).  Even though our lowest-mass
and highest-mass galaxies in SIDM-Hydro differ by a factor of 10 in stellar
mass, their profiles show much smaller differences with respect to their
SIDM-DMO runs in contrast to the CDM-hydro vs CDM-DMO results.

Figure~\ref{fig:figure4} highlights the differences between CDM and SIDM as a
function of galaxy stellar mass by showing only the hydro density profiles
(solid for CDM, dotted for SIDM). In the left panels, we plot the density
profiles for the three lowest stellar mass systems, while the right panel shows
the density profile of the highest stellar mass galaxy. Only the galaxy with the
highest stellar mass ($\mstar = 10^7\,\msun$, right panel) forms a core in the CDM runs,
while all galaxies have sizable cores in the SIDM versions. In all three of the
lower stellar mass systems, the central dark matter properties are primarily
determined by the dark matter physics, with baryonic effects playing a minimal
role. It is only in the highest $\mstar$ galaxy that baryons significantly alter
the structure in the CDM halo (and further reduce the density in the SIDM). This
result further strengthens the picture in which galaxies with $\mstar \la
3\times10^6\,\msun$ have dark matter density profiles that are essentially
unmodified by baryons \citep{ono15,chan15,tollet16}.

\subsection{Density profile slopes}
\label{subsec:slopes}
Results in the previous subsection demonstrate that feedback can reduce the
central dark matter density in CDM halos, provided enough stars form.  The same
subsection also demonstrates that SIDM alone can do so as well.  However, the
precise nature of this reduction is important, and in this subsection, we study
the slopes of the density profiles quantitatively.  We obtain the inner slope of
the DM density assuming a power law and apply the $\chi ^2$ fitting method to
the density profiles within 250-500 pc range, which is comparable to 0.5-1\% the
virial radii of their DM haloes.

Figure~\ref{fig:figure6} shows the resulting logarithmic slope
$\alpha(r)=\rm{d}\log \rho/\rm{d}\log \textit{r}$ of the hydro (filled symbols)
and DMO (empty symbols) simulations as a function of the stellar mass of the
galaxy (top row). Also shown is the slope of the profiles versus radial distance
from the halo centers (bottom row), with arrows marking the stellar half-mass
radii for each halo. For the estimation of the central slope, we varied the
fitting range and the bin size and found slopes that do not differ by more than
0.1 dex; this uncertainty is accounted for by the size of the symbols in the
figure.  We find that only one of the CDM-Hydro simulations in our sample truly
becomes ``cored'' (defined here as $\alpha > -0.3$), and even then, this happens
only at very small radii ($r \la 300$ pc). As the stellar mass of the galaxies
decreases, the inner slopes in the CDM-Hydro simulations decrease to the
mild-cusp $-0.6\leq\alpha<-0.3$ and to the cuspy region ($\alpha<-0.6$). The
cuspy inner slopes in the CDM-DMO runs remain largely unaffected by stellar
feedback from FIRE for galaxies with $\log \mstar / \msun < 6.2$ and have only a
mild change for the galaxy with $\log \mstar / \msun \sim 6.6$.

In contrast, all SIDM simulations (DMO and Hydro) exhibit central density cores.
Despite varying in an order of magnitude in $\mstar$, the SIDM-Hydro simulations
all have central density profiles with slopes of $\alpha > -0.5$.  More
importantly, the slopes in the hydro runs closely follow their DMO values, even
for the highest stellar masses. The close similarity between the density
profiles of the SIDM-DMO and SIDM-Hydro runs -- including the similar shape of
$\alpha(r)$ across all values of $\mstar$ -- indicates that independently of the
galaxy mass and SFH, core formation and reduction of central densities in SIDM
simulations are set mainly by dark matter physics rather than by galaxy
formation physics (for the cross section $\sigma /m=1 \, \cmg$ studied here).
This provides a striking contrast to the major role that feedback plays in
forming cores in CDM simulations.  In fact, the stellar mass dependence of the
density profile slope in CDM-Hydro simulations is seen at radii of up to $\sim$
1 kpc.  \textit{SIDM predictions regarding the central gravitational potential
  of $\mstar \sim 10^{6} \, M_{\odot}$ dwarf galaxies appear relatively robust
  to the effects of stellar feedback, while CDM predictions depend sensitively
  on it.}

The changes in DM densities found in SIDM-Hydro versus CDM-Hydro simulations are
quantified in more detail in Figure~\ref{fig:figure7}, which shows the
difference in $\alpha(r)$ between these runs. For the same FIRE physics, the
SIDM densities are more than 25\% different from the CDM densities for $r < 500$
pc (and can be over 50\% different at $250\,{\rm pc}$).  This ratio shows little
dependence on $\mstar$.  The left panel shows that less massive galaxies exhibit
larger differences in the slope, with the largest change happening within the
SIDM half-mass radii (where self-interactions form the core). The smallest
difference in the slope occurs for the most massive galaxy, as feedback in the
CDM version of this halo is strong enough to create a core similar to its SIDM
analog.

\subsection{Shapes}
\label{subsec:shapes}
\begin{figure}
	\includegraphics[width=\columnwidth]{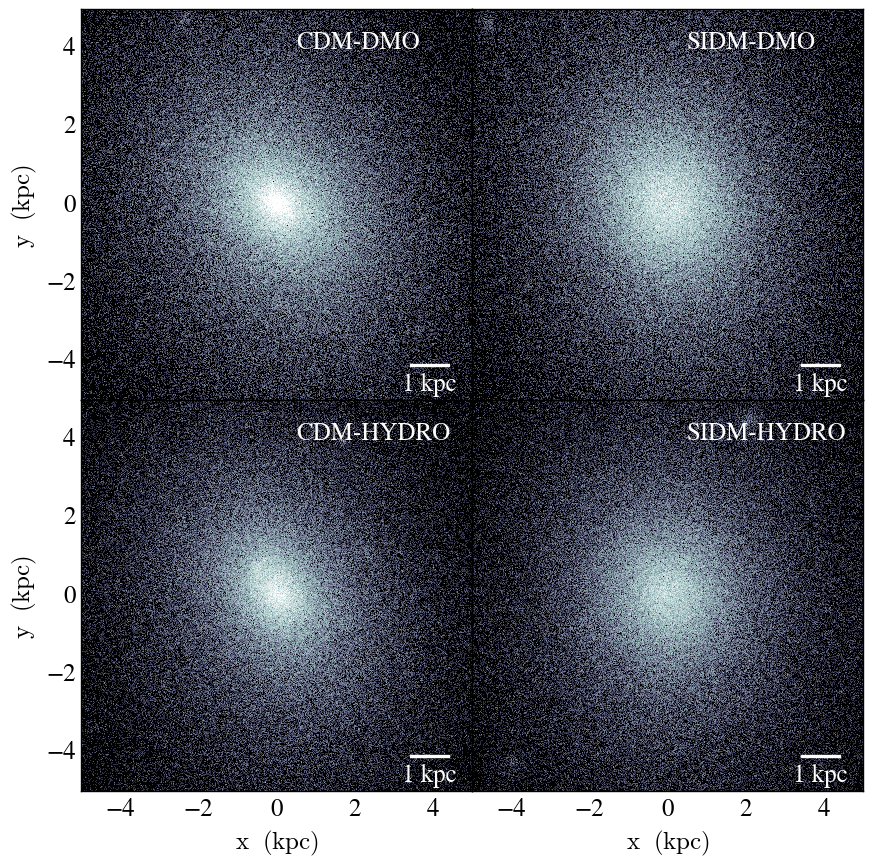}
        \caption{Projected ($x-y$ plane) visualization of the dark matter within
          a $5\,$ kpc radius from the center for m10f
          ($M_{\star,\mathrm{cdm}} \approx 4.11 \times \,10^6\msun $). The
          galaxy ends with roughly equal stellar mass in both DM models, the
          FIRE-2 baryonic physics reduces the central dark matter density for
          the CDM-Hydro simulation compared to the DMO, whereas the same
          feedback physics has a milder effect in the SIDM-Hydro simulation.}
\label{fig:figure5}
\end{figure}

Many studies have shown that CDM haloes in DMO simulations are triaxial
\citep{vega17,schneider12,sprin04}. SIDM haloes are expected to be closer to
spherical in the region for which self-interactions are important, as the
interactions tend to isotropize the density distribution
\citep{spe00,kap14,rocha13,peter13,zav13,elb15}. The shapes of low-mass dark
matter haloes and their dwarf galaxies may therefore contain important clues
about the nature of dark matter.

We show a visualization of the DM distribution corresponding to one of our
simulations (m10d in Table \ref{tab:table1}) in Figure~\ref{fig:figure5}. The
SIDM-DMO run indeed exhibits a distinctive roundness within the half-mass radius
($\sim 1\,{\rm kpc}$), while the CDM-DMO run is noticeably more triaxial. In
both models, the inclusion of hydrodynamics mildly affects the DMO predictions.

To characterize halo shapes, we compute shape tensors using an iterative method
\citep{dub91, zemp11}. The shape tensor eigenvalues are proportional to the
square root of the principal axes of the ellipsoid that characterize the
particle distribution. Following the standard nomenclature for the
semi-principal axes $a$, $b$ and $c$, we choose $a \geq b \geq c$ and calculate
the axis ratios $b/a$ and $c/a$.  In general, $c/a$ and $b/a$ quantify the
degree of triaxiality of the distribution under study, because $c$ is the
smallest of the semi-principal axes, then $c/a \approx 1$ will imply $b/a$ is
also close to unity and any deviations from spherical symmetry will be small.

Table ~\ref{tab:table2} summarizes the axis ratios $b/a$ and $c/a$ for the
central DM distribution using the particles within $1$ kpc for each of the
haloes. For the hydrodynamical simulations, we also include the axis ratios of
the stars within the same radius. We can assess the effect of the dark matter
properties and/or the feedback on the shape of the inner DM mass distribution in
dwarf galaxies\footnote{It is important to note that we are measuring the
  triaxial distribution for particles within $1$ kpc where the feedback has the
  largest effect on the DM; the results are unchanged if we consider the axis
  ratios at $0.5$ kpc or $1.5$ kpc rather than at $1$ kpc. The values presented
  in Table~\ref{tab:table1} are only meant to characterize the shape of the
  inner region of the halo, not its entire extent.} by computing the 3D-axis
ratios at the typical size of the visible matter in dwarf galaxies for the Hydro
and DMO runs in both DM models.

\begin{table*}
  \caption{3D-axis ratio between the smallest and the largest
    semi-principal axis (c/a) of the DM and of the stellar mass
    distributions computed at 1 kpc. Columns: (1) Halo name used in the suite of
    \citet{fitts16}; (2)-(9) DM axis ratios (b/a and c/a) for each of the 4
    simulations for both cases, DMO and DM+Hydro, as indicated by the column
    labels. Columns (10)-(13) show the 3D-axis ratios (b/a and c/a) for the
    stellar component in CDM and SIDM, respectively, also calculated at 1 kpc.}
	\label{tab:table2}
	\resizebox{\textwidth}{!}{
	\begin{tabular}{lcccccccccccr} 
		\hline
		 & \multicolumn{4}{c}{Dark Matter-Only (DMO)} & \multicolumn{4}{c}{DM+Hydro}& \multicolumn{4}{c}{Stars} \\	
		 \hline
		Halo	& \multicolumn{2}{c}{CDM} & \multicolumn{2}{c}{SIDM} & \multicolumn{2}{c}{CDM} & \multicolumn{2}{c}{SIDM} &  \multicolumn{2}{c}{CDM} & \multicolumn{2}{c}{SIDM} \\
		  & b/a & c/a & b/a & c/a & b/a & c/a &b/a & c/a &b/a & c/a  &b/a & c/a\\		
		\hline
		m10b & 0.50 & 0.38  & 0.90 &  0.76 & 0.55 & 0.43 & 0.80 & 0.68  & 0.58 &0.49 & 0.70 & 0.43  \\ 
		m10d &  0.57 & 0.44  & 0.87 & 0.84 & 0.56 & 0.44 & 0.85 & 0.76 & 0.63 & 0.49 & 0.81 & 0.65 \\ 
		m10f & 0.52 & 0.42 &  0.95  & 0.91 & 0.57  & 0.43 & 0.88  & 0.79 & 0.60 & 0.46 & 0.69 & 0.56 \\ 
		m10k & 0.55 & 0.41  & 0.94 & 0.85 & 0.60 & 0.45 & 0.81 & 0.72 & 0.60 & 0.45 & 0.67 & 0.51 \\ 
		\hline
	\end{tabular}
	}
\end{table*}

We find a systematic preference for the cuspy CDM haloes (both DMO and Hydro) to
be triaxial: even the galaxy with a core (m10k) is less round (lower $c/a$
ratio) than any of the SIDM haloes. While the galaxy formation physics in the
FIRE-2 model affects the inner shapes of the halos in SIDM-Hydro runs, those
halos remain rounder than the versions in the CDM-Hydro runs. The galaxies
formed in both cases are fairly triaxial, though the SIDM galaxies are slightly
closer to spherical. Galaxy formation physics (as opposed to gravitational
physics or self-interactions) therefore appears to play the dominant role in
establishing shapes of dwarf galaxies in these simulations.

\section{Discussion and Conclusions}
\label{sec:conclusion}
SIDM preserves the successes of $\Lambda$CDM on large scales while
simultaneously providing a path to ameliorate small-scale challenges to the
model \citep{bull17}. The main effect of SIDM on dark matter haloes is to reduce
the density and sphericalize the dark matter distribution on scales where many
dark matter self-interactions can occur per Hubble time
\citep{elb15,rocha13,kap14,kap16}. In order to understand \textit{observable}
consequences of SIDM, however, we must study the combined effects of SIDM and
galaxy formation physics.

In this paper, we present high-resolution SIDM cosmological simulations (with
$\sigma/m=1\,\cmg$) of four isolated dwarf galaxies taken from a large suite of
$\mhalo(z=0) \approx 10^{10}\,\msun$ haloes \citep{fitts16}. In each case, we
have dark-matter-only and hydrodynamical simulations; the hydrodynamical
simulations employ an identical model of galaxy formation physics (FIRE-2) to
the CDM versions of the haloes presented in \citet{fitts16}. Accordingly, we are
able to understand the modification of halo properties due to dark matter
self-interactions alone (by comparing CDM-DMO and SIDM-DMO runs) and
modifications coming from a combination of dark matter physics and galaxy
formation physics (by comparing both hydro runs).  The high spatial and mass
resolution of our simulations allow us to unambiguously address the impact of
stellar feedback on the core formation and density reduction within 1 kpc of
each of the SIDM and CDM galaxies.

We focus on the comparison of DM profiles in DMO and hydro simulations for the
SIDM and CDM models.  We show that SIDM galaxies display similar star formation
histories as their CDM counterparts, resulting in nearly identical stellar
masses and sizes in each case. The sole exception is the lowest-mass galaxy,
which forms twice as many stars in SIDM but it nonetheless follows the same
stellar mass - size relation as rest of the sample, which is essentially
identical in CDM and SIDM (see Figure \ref{fig:figure2}).

In the CDM simulations, the main mechanism to modify a central dark matter cusp is stellar
feedback. As demonstrated by \citet{fitts16}, the effects of stellar feedback
at the halo mass scale considered here -- $10^{10}\,\msun$ -- are strongly
dependent on stellar mass (see also \citet{chan15,tollet16,ono15}). Galaxies with $\mstar \la
3\times 10^6\,\msun$ maintain the central cusp found in DMO runs, while those
with $\mstar \ga 3\times 10^6\,\msun$ have reduced central densities, with the
reduction increasing with stellar mass. SIDM produces qualitatively different
results: the central densities in DMO simulations are reduced significantly
through dark matter self-interactions. When considering the change between
DMO and hydro runs in SIDM, however, differences are minimal:
the dark matter core sizes and density profiles in the full physics runs are
generically very similar to their DMO counterparts. \textit{Feedback only has a minimal
effect on the dark matter structure of SIDM dwarf galaxies over the mass range
simulated here ($10^6 \la \mstar/\msun \la 10^7$).}

Based on our results, the discovery of dark matter cores on the scale of
$r_{1/2}$ in field dwarf galaxies with $\mstar \la 3 \times 10^6\,\msun$ would
imply one of the following: (1) dark matter is cold but the implementation of
astrophysical processes in current codes is incomplete; (2) there is a large
scatter in the halo masses of dwarf galaxies with
$\mstar \la 3 \times 10^6\,\msun$; or (3) dark matter has physics beyond that of
a cold and collisionless thermal relic -- perhaps self-interactions of the kind
explored here. The shape of the dark matter density profiles in
$\mstar \sim 10^6\,\msun$ isolated dwarf galaxies on scales comparable to the
galaxy half-mass radius therefore provide a crucial test of dark matter models.

\section*{Acknowledgements}

V.H.R. acknowledges support from UC-MEXUS and CONACyT through the postdoctoral
fellowship. A.G-S. acknowledges support from UC-MEXUS through the postdoctoral Fellowship. 
JSB is supported by NSF and HST grants at UC Irvine.  MBK
acknowledges support from NSF grant AST-1517226 and from NASA grants NNX17AG29G
and HST-AR-12836, HST-AR-13888, HST-AR-13896, and HST-AR-14282 from the Space
Telescope Science Institute, which is operated by AURA, Inc., under NASA
contract NAS5-26555. DK was supported by NSF grant AST-1412153 and the Cottrell
Scholar Award from the Research Corporation for Science Advancement.  CAFG was
supported by NSF through grants AST-1412836 and AST-1517491, and by NASA through
grant NNX15AB22G. Our simulations used computational resources provided via the
NASA Advanced Supercomputing (NAS) Division and the NASA Center for Climate
Simulation (NCCS) and the Extreme Science and Engineering Discovery Environment,
which is supported by National Science Foundation grant number OCI-1053575.

\bibliographystyle{mnras}
\bibliography{bib_reference_list} 


\bsp	
\label{lastpage}
\end{document}